\documentclass{article}

\usepackage{algorithm}
\usepackage{algorithmic}
\usepackage{listings}
\usepackage{graphicx}
\usepackage{subfigure}
\usepackage[left=2.5cm,right=2.5cm,bottom=2.5cm,top=2.5cm]{geometry}
\usepackage{array,booktabs}
\newcolumntype{L}{@{}>{\kern\tabcolsep}l<{\kern\tabcolsep}}
\usepackage{colortbl}
\usepackage{xcolor}
\usepackage{caption}
\usepackage{color}
\usepackage{amsmath,amsfonts,amssymb}
\usepackage[utf8]{inputenc}
\usepackage[english]{babel}
\usepackage[nottoc]{tocbibind}
\usepackage[fleqn,tbtags]{mathtools}
\usepackage{stmaryrd}
\usepackage{gensymb}
\usepackage[title]{appendix}

\newcommand{\op}{\overline{p}}
\newcommand{\ot}{\overline{\theta}}
\newcommand{\on}{\overline{n}}
\newcommand{\oepsilon}{\overline{\varepsilon}}
\newcommand{\ovr}{\overline{r}}
\newcommand{\oN}{\overline{N}}
\newcommand{\oJ}{\overline{J}}
\newcommand{\ophi}{\overline{\phi}}
\newcommand{\obeta}{\overline{\beta}}
\newcommand{\oG}{\overline{G}}
\newcommand{\oB}{\overline{B}}

\newcommand{\oW}{\overline{W}}

\newcolumntype{C}[1]{>{\centering\arraybackslash}p{#1}}

\DeclareMathOperator*{\argmin}{\arg\min}

\begin{document}
	
	\title{Photovoltaic effect in multi-domain ferroelectric perovskite oxides}
	\author{Ying Shi Teh and Kaushik Bhattacharya\\ \small Division of Engineering and Applied Science\\ \small California Institute of Technology\\ \small Pasadena, CA 91125 USA}
	\date{}
	
	\maketitle

\begin{abstract}
	We propose a device model that elucidates the role of domain walls in the photovoltaic effect in multi-domain ferroelectric perovskites.  The model accounts for the intricate interplay between ferroelectric polarization, space charges, photo-generation and electronic transport.  When applied to bismuth ferrite, results show a significant electric potential step across both $71\degree$ and $109\degree$ domain walls, which in turn contributes to the photovoltaic (PV) effect. We also find a strong correlation between polarization and oxygen octahedra tilts, which indicates the nontrivial role of the latter in the PV effect. The domain wall-based PV effect is further shown to be additive in nature, allowing for the possibility of generating above-bandgap voltage.
\end{abstract}

%===================================================================================
\section{Introduction}
%===================================================================================

In conventional photovoltaics, electron-hole pairs are created by the absorption of photons that are then separated by an internal field in the form of heterogeneous junctions such as p-n junctions. Less than a decade ago, Yang {\it et al.}  \cite{Yang2010} reported large photovoltages generated in thin films of multi-domain bismuth ferrite (BFO), and suggested a new mechanism where electrostatic potential steps across the ferroelectric domain walls drives the photocurrent. This discovery has since revitalized the field of photoferroics. Among many ferroelectric oxides, BFO has particularly attracted considerable interest due to its high ferroelectric polarization and relatively small bandgap. 

Many novel experiments were subsequently devised to investigate the role of domain walls in the observed photovoltaic (PV) effect in ferroelectric perovskites. Alexe and Hesse  \cite{Alexe2011} performed measurements of the local photoelectric effect using atomic force microscopy (AFM). They found that the photocurrent is essentially constant across the entire scanned area, hence indicating the absence of the domain wall (DW) effect. The nanoscale mapping of generation and recombination lifetime using a method combining  photoinduced transient spectroscopy (PITS) with scanning probe microscopy (SPM) points to a similar conclusion \cite{Alexe2012}.  This led to the hypothesis that the bulk photovoltaic (BPV) effect, which arises from the noncentrosymmetry of perovskites, is the key mechanism instead \cite{Bhatnagar2013,Yang2017}. However further recent studies focused on characterizing both the BPV and DW effects show that the latter effect is much more dominant \cite{Inoue2015,Matsuo2016}. The lack of clear concensus among the scientific community on the key mechanism in the PV effect in perovskites as well as on the role of domain walls could be understood from the inherent difficulties in the experimental techniques. The nanoscale order of ferroelectric domain walls makes it difficult to probe into and separate the effects from the bulk domains and the domain walls. Other issues such as defect formation and grain boundaries in perovskite crystals further complicate the analysis.

First-principles calculations have provided a detailed understanding of the structure of domain walls \cite{Meyer2002,Lubk2009,Dieguez2013}, and have established the drop in electrostatic potential across it.  However, they are limited to a few nanometers, and cannot examine the interaction of domain walls with other features.  On the other hand, models at the device scale provide understanding at the scale \cite{Seidel2011}, but assume {\it a priori} the polarization and other aspects of the domain wall.  Finally, various phase field models provide understanding of the domain pattern \cite{Xue2014,Cao2016}, but in the absence of space charge and photocurrent. Thus, there is a gap in our modeling of the  intricate interplay between space charge, ferroelectric polarization and electronic transport.

This paper seeks to fill this gap by building on prior work of Xiao \textit{et al.} \cite{Xiao2005,Xiao2008} and Suryanarayana and Bhattacharya \cite{Suryanarayana2012} who developed a continuum theory of semiconducting ferroelectrics including electron and hole transport.  We extend their work to include photogeneration due to illumination and study photovoltaic effect in ferroelectric perovskite oxides. We investigate the photovoltaic response of BFO films with different domain wall configurations by solving the model at the device scale. At a 71$\degree$ or 109$\degree$ domain wall, we observe a change in the component of polarization perpendicular to the domain wall. This in turn results in a relatively large electrostatic potential step across the wall which allows for separation of photogenerated electron-hole pairs.  Thus this model supports the hypothesis of domain walls contributing to the photovoltaic effect.

We emphasize that this model does not {\it a priori} assume the domain wall structure or the electrostatic potential step across it.  Instead, this is a prediction of the model that is based on well-established Devonshire-Landau models of ferroelectrics and lumped band models of semiconductors.

The rest of the paper is organized as follows:  In Section \ref{Section: BFO structure}, we briefly review the structure of BFO and discuss the classifications of ferroelectric domain walls in BFO.  We develop the theoretical framework in Section \ref{Section:Theory}.  We apply the theory to examine PV effect in ferroelectrics with domain walls in Section \ref{Section:App}.  We conclude with a brief discussion in Section \ref{Section:Conc}.

%===================================================================================
\section{Bismuth Ferrite} \label{Section: BFO structure}
%===================================================================================

In this work, we focus on bismuth ferrite (BFO), though we note that the same framework can be applied to other ferroelectrics and the results are expected to be similar qualitatively.

\begin{figure}
	\begin{center}
		\includegraphics[width=0.3 \textwidth]{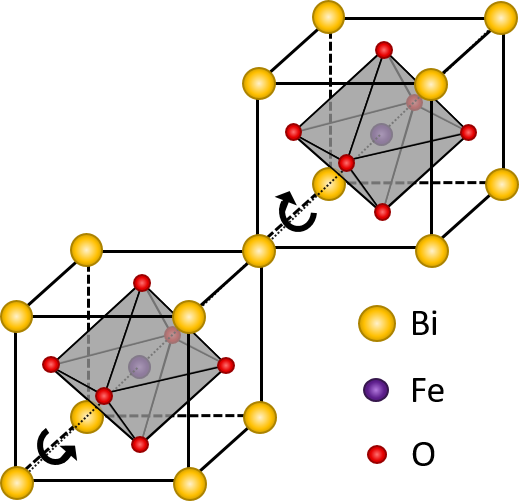}
	\end{center}
	\caption{Crystal structure of bulk BFO. The two O$_6$ octahedra rotate out-of-phase about the polarization axis marked by the dotted line.}
	\label{Fig: BFO}
\end{figure}

At room temperature, BFO has a rhombohedral phase with space group $R3c$ (Figure \ref{Fig: BFO}). The displacements of the atoms from the ideal cubic structure in this phase lead to a spontaneous polarization pointing in the [111] pseudocubic direction. Another distintive feature is the network of O$_6$ octahedra surrounding the Fe ions that rotate or tilt out-of-phase about the polarization axis. This is also commonly known as the antiferrodistortive (AFD) mode and has been found to play an important role in the ferroelectric phase of the material \cite{Dieguez2013}.

Electric polarization in rhombohedral BFO can take one of the eight variants of the [111] pseudocubic direction which gives possible domain wall orientations of 71$\degree$, 109$\degree$ and 180$\degree$. On each domain, there can be two possible orientations of oxygen octahedra. We follow Lubk \textit{et al.}'s method of classifying oxygen octahedra tilts (OTs) across domain walls as either \textit{continuous} or \textit{discontinuous} \cite{Lubk2009}. In the continuous case, the direction of oxygen octahedra tilt remains the same along the polarization vector field. In the discontinuous case, the direction reverses across the domain wall. 

%===================================================================================
\section{Theory} \label{Section:Theory}
%===================================================================================

We consider a metal-perovskite-metal (MPM) structure that is connected to an external voltage source to form a closed electrical circuit (see Figure 1). The multi-domain ferroelectric perovskite film occupying the space $\Omega$ is subjected to light illumination. The two metal-perovskite interfaces are denoted by $\partial \Omega_1$, $\partial \Omega_2$ $\in \partial \Omega$. All the processes are assumed to occur at constant temperature $T$.  We present the equations and their physical meanings here. Readers may refer to Appendix \ref{Appendix: dissipation rate} for the thermodynamically consistent derivation.

\begin{figure}
	\begin{center}
		\includegraphics[width=0.3 \textwidth]{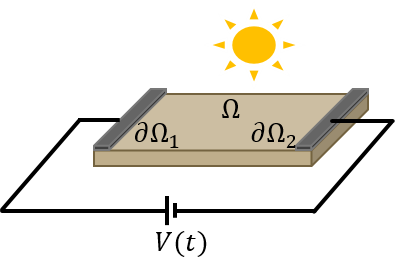}
	\end{center}
	\caption{Schematic of a device model in a metal-perovskite-metal configuration.}
\end{figure}

%-----------------------------------------------------------------------------------
\subsection{Charge and electrostatic potential}
%-----------------------------------------------------------------------------------
% Gauss' law for electricity

The total charge density ($\mathbf{x} \in \Omega$) is given by 
\begin{equation}
	\rho = q(p_v - n_c + z_d N_d^+ - z_a N_a^-),
\end{equation}
where q is the electronic charge, $n_c$ is the density of electrons in the conduction band, $p_v$ is the density of holes in the valence band, $N_d^+$ is the density of ionized donors, $N_a^-$ is the density of ionized acceptors, $z_d$ is the valency number of donors, and $z_a$ is the valency number of acceptors.
The polarization and space charge in the ferroelectrics together generate an electrostatic potential. This is determined by Gauss' equation 
\begin{equation}\label{Eqn: Gauss}
	\nabla \cdot (-\varepsilon_0 \nabla \phi + \mathbf{p}) =  \rho,
\end{equation}
where $\varepsilon_0$ is the permittivity of free space, subject to appropriate boundary conditions.

%-----------------------------------------------------------------------------------
\subsection{Transport equations}
%-----------------------------------------------------------------------------------

In the presence of light illumination, an incident photon may be absorbed in the semiconductor to promote an electron from the valence band to the conduction band, thus generating an electron-hole pair in the process of photogeneration. The reverse may also occur such that an electron and a hole recombine. Electrons and holes may also move from one point to another point, as represented by the electron and hole density flux terms, $\mathbf{J}_n$ and $\mathbf{J}_p$. With conservation of electrons and holes, we can relate the time derivatives of densities of electrons and holes to the aforementioned processes via the following transport equations,
\begin{equation}\label{Eqn: Transport nc}
	\dot{n_c} = -\nabla \cdot \mathbf{J}_n + G - R,
\end{equation}
\begin{equation}\label{Eqn: Transport pv}
	\dot{p_v} = -\nabla \cdot \mathbf{J}_p + G - R,
\end{equation}
where $G$ is the rate of photogeneration which can be taken to be proportional to the intensity of light illumination, while $R$ is the recombination rate. Here we assume that the only form of recombination present is radiative recombination, which involves the transition of an electron from the conduction band to the valence band along with the emission of a photon. $R$ takes the form of $ B(n_c p_v - N_i^2) $ with the intrinsic carrier density being given by  $N_i =\sqrt{N_c N_v} \exp ( -\frac{E_c - E_v}{2k_B T} )$ \cite{Nelson2003}. The radiative recombination coefficient $B$ is a material property, independent of the carrier density.

The electron and hole fluxes  $\mathbf{J}_n,  \mathbf{J}_p$ are taken to be proportional to the gradient in its electro-chemical potential \cite{Nelson2003} via
\begin{equation}\label{Eqn: Jn}
\mathbf{J}_n = - \frac{1}{q} \nu_n n_c \nabla \mu_n,
\end{equation}
\begin{equation}\label{Eqn: Jp}
\mathbf{J}_p = - \frac{1}{q} \nu_p p_v \nabla \mu_p,
\end{equation}
where $\nu_n$ and $\nu_p$ are the electron and hole mobilities respectively.

In this work, the diffusion of donors and acceptors are neglected.

%-----------------------------------------------------------------------------------
\subsection{Free energy}
%-----------------------------------------------------------------------------------

The free energy of the ferroelectric is postulated to be of the form
\begin{equation}\label{Eqn: W}
	W  = W_{DL}(\mathbf{p},\mathbf{\theta}) + W_G(\nabla\mathbf{p}, \nabla\mathbf{\theta}) + W_{n_c}(n_c) + W_{p_v}(p_v) + W_{N_d}(N_d^+) + W_{N_a}(N_a^-).
\end{equation}
The various terms are currently explained. 

$W_{DL}$ refers to the Devonshire-Landau free energy of bulk ferroelectrics. In addition to the typical primary order parameter of electric polarization $\mathbf{p}$, we include a second order parameter --- oxygen octahedral tilts $\mathbf{\theta}$. We adopt the following energy form for BFO \cite{Xue2014}. The corresponding coefficients can be found in Table \ref{Table: LGD parameters}.
\begin{equation}\label{Eqn: GLD energy}
\begin{split}
W_{DL} = & a_1(p_1^2 + p_2^2 + p_3^2) + a_{11}(p_1^4 + p_2^4 + p_3^4) + a_{12}(p_1^2 p_2^2 + p_2^2  p_3^2 + p_1^2 p_3^2)\\
& + b_1(\theta_1^2 + \theta_2^2 + \theta_3^2) + b_{11}(\theta_1^4 + \theta_2^4 + \theta_3^4) + b_{12}(\theta_1^2 \theta_2^2 + \theta_2^2 \theta_3^2 + \theta_1^2 \theta_3^2) \\
&+ c_{11}(p_1^2 \theta_1^2 + p_2^2 \theta_2^2 + p_3^2 \theta_3^2)
+ c_{12}[p_1^2(\theta_2^2 + \theta_3^2) + p_2^2(\theta_1^2 + \theta_3^2) + p_3^2(\theta_1^2 + \theta_2^2)] \\
&+ c_{44} (p_1 p_2 \theta_1 \theta_2 + p_1 p_3 \theta_1 \theta_3 + p_2 p_3 \theta_2 \theta_3).
\end{split}
\end{equation} 

The energy stored in the ferroelectric domain walls is accounted for through the gradient or Ginzburg energy term $W_G$ which includes the energy cost associated with rapid change in polarization and octahedral tilts.  
\begin{equation}\label{Eqn: Gradient energy}
	W_G(\nabla\mathbf{p}, \nabla\mathbf{\theta}) = \frac{1}{2}a_0 |\nabla \mathbf{p}|^2 + \frac{1}{2}b_0 |\nabla \mathbf{\theta}|^2.
\end{equation}
Here we assume that the gradient terms are isotropic for simplicity but can easily be modified. 

$W_{n_c}, W_{p_v}, W_{N_d}, W_{N_a}$ in equation (\ref{Eqn: W}) are the free energies of electrons in the conduction band, holes in the valence band, donors and acceptors respectively. The explicit expressions of these energies can be determined by considering each system as a canonical ensemble in the framework of statistical mechanics\cite{Suryanarayana2012}
\begin{equation}\label{Eqn: Wnc}
	W_{n_c}(n_c) = n_c E_c + k_B T [-N_c \log N_c + n_c \log n_c + (N_c - n_c)\log(N_c - n_c)],
\end{equation} 
\begin{equation}
	W_{p_v}(p_v) = (N_v - p_v) E_v + k_B T [-N_v \log N_v + p_v \log p_v + (N_v - p_v)\log(N_v - p_v)],
\end{equation} 
\begin{equation}
\begin{split}
	W_{N_d}(N_d^+) = &(N_d - N_d^+) E_d - (N_d - N_d^+)k_B T \log(2z_d)\\& + k_B T [-N_d \log N_d + N_d^+ \log N_d^+ + (N_d - N_d^+)\log(N_d - N_d^+)],
\end{split}
\end{equation} 
\begin{equation}\label{Eqn: WNa}
\begin{split}
	W_{N_a}(N_a^-) = & N_a^- E_a - (N_a - N_a^-)k_B T \log(2z_a) \\& + k_B T [-N_a \log N_a + N_a^- \log N_a^- + (N_a - N_a^-)\log(N_a - N_a^-)].
\end{split}
\end{equation} 

%-----------------------------------------------------------------------------------
\subsection{Polarization and tilt equations} \label{Section: PT}
%-----------------------------------------------------------------------------------

The polarization and tilt evolve according to the (time-dependent) Landau-Ginzburg equations
\begin{equation}\label{Eqn: p}
\frac{1}{\nu_{\mathbf p}} \dot{\mathbf p} = \nabla \cdot \frac{\partial W_G}{\partial \nabla \mathbf{p}} - \frac{\partial W_{DL}}{\partial \mathbf{p}} - \nabla \phi,
\end{equation}
\begin{equation}\label{Eqn: theta}
\frac{1}{\nu_{\mathbf \theta}} \dot{\mathbf \theta} = \nabla \cdot \frac{\partial W_G}{\partial \nabla \mathbf{\theta}} - \frac{W_{DL}}{\partial \mathbf{\theta}}.
\end{equation}
where ${\nu_{\mathbf p}}, {\nu_{\mathbf \theta}}$ are the respective mobilities.  They are subject to natural boundary conditions
\begin{equation}
\mathbf{\hat{n}} \cdot \frac{\partial W_g}{\partial \nabla \mathbf{p}} = 0,
\end{equation}
\begin{equation}
\mathbf{\hat{n}} \cdot \frac{\partial W_g}{\partial \nabla \mathbf{\theta}} = 0.
\end{equation}

%-----------------------------------------------------------------------------------
\subsection{Electrochemical potentials} \label{Section: EC pot}
%-----------------------------------------------------------------------------------

The electrochemical potentials are obtained from the energy to be 
\begin{equation}
\mu_n = \frac{\partial W_{n_c}}{\partial n_c} - q\phi,
\end{equation}
\begin{equation}
\mu_p = \frac{\partial W_{p_v}}{\partial p_v} + q\phi,
\end{equation}
\begin{equation}
\mu_{N_d^+} = \frac{\partial W_{N_d}}{\partial N_d^+} + qz_d\phi,
\end{equation}
\begin{equation}\label{Eqn: muNa}
\mu_{N_a^-} = \frac{\partial W_{N_a}}{\partial N_a^-} - qz_a\phi.
\end{equation}
At thermal equilibrium, $\mu_n = -\mu_p = - \mu_{N_d^+} = \mu_{N_a^-} = E_{F_m} $ where the magnitude of $E_{F_m}$ is the workfunction of the metal electrode.  Further, using equations (\ref{Eqn: Wnc}) to (\ref{Eqn: WNa}) we can invert the relations to obtain
\begin{equation}
\begin{split}
& \qquad \qquad \ n_c = \frac{N_c}{1 + \exp (\frac{E_c - E_{F_m} - q\phi}{k_B T})}, \\
& \qquad \qquad \ p_v = \frac{N_v}{1 + \exp (\frac{E_{F_m - E_v + q\phi}}{k_B T})}, \\
& \quad N_d^+ = N_d \bigg[ 1 - \frac{1}{1 + \frac{1}{2z_d} \exp(\frac{-E_{F_m} + E_d - q\phi z_d}{k_B T})} \bigg], \\
& N_a^- = N_a \bigg[ 1 + 2z_a \exp (\frac{-E_{F_m} + E_a - q \phi z_a}{k_B T}) \bigg]^{-1}, \\
\end{split}
\end{equation}
consistent with Fermi-dirac distribution \cite{Sze1969}.  Finally, assuming $N_c >> n_c$ and $N_v >> p_v$, equations (\ref{Eqn: Jn}) and (\ref{Eqn: Jp}) become

\begin{equation}\label{Eqn: Jn2}
\mathbf{J}_n = - \frac{\nu_n k_B T}{q} \nabla n_c + \nu_n n_c \nabla \phi,
\end{equation}
\begin{equation}\label{Eqn: Jp2}
\mathbf{J}_p = - \frac{\nu_p k_B T}{q} \nabla p_v - \nu_p p_v \nabla \phi.
\end{equation}
Equations (\ref{Eqn: Jn2}) and (\ref{Eqn: Jp2}) show that each of $\mathbf{J}_n$ and $\mathbf{J}_p$ can be resolved into two contributions: (1) a diffusion current, driven by concentration gradient of carriers, and (2) a drift current, driven by an electric field. By applying the Einstein relation which relates diffusion constant $D$ to mobility $\nu$ via $D = \nu k_B T / q$, we recover the equations that are typically written to describe the flow of electrons and holes in solar cells. 

%-----------------------------------------------------------------------------------
\subsection{Ohmic boundary conditions} \label{Section: Ohmic BCs}
%-----------------------------------------------------------------------------------

We prescribe ohmic boundary conditions at the contacts with metal electrodes following \cite{Foster2014} for convenience.  We have
\begin{equation*}
	\begin{rcases}
		n_c = N_c e^{-(E_c - E_{f_m}) / k_B T} \quad \\
		p_v = N_v e^{-(E_{f_m} - E_v) / k_B T} \quad
	\end{rcases}
	\quad \text{on } \partial \Omega_1 \cup \partial \Omega_2.
\end{equation*}
This is equivalent to assuming that the Fermi level in the semiconductor aligns with that of the metal, hence giving rise to electron and hole densities that are independent of the applied voltage.

%-----------------------------------------------------------------------------------
\subsection{Steady-state model}
%-----------------------------------------------------------------------------------

At steady state, all the fields of interest do not vary with respect to time. Further, we are interested in domain walls, and therefore can assume that things are invariant parallel to the domain wall.  This means that we have one independent space variable which we denote $r$.  We denote the components of polarization and tilt parallel (respectively perpendicular) to the domain wall to be $p_s, \theta_s$ (respectively $p_r, \theta_r$). With $z_d = z_a = 1$, we have a coupled system of differential equations for region $x \in (0,L)$, where $L$ is the length of the film.
\begin{equation}
	a_0 \frac{d^2p_r}{dr^2} - \frac{\partial W_{DL}}{\partial p_r} - \frac{d\phi}{dr} = 0,
\end{equation}
\begin{equation}
	a_0 \frac{d^2p_s}{dr^2} - \frac{\partial W_{DL}}{\partial p_s} = 0,
\end{equation}
\begin{equation}
	b_0 \frac{d^2\theta_r}{dr^2} - \frac{\partial W_{DL}}{\partial \theta_r} = 0,
\end{equation}
\begin{equation}
	b_0 \frac{d^2\theta_s}{dr^2} - \frac{\partial W_{DL}}{\partial \theta_s} = 0,
\end{equation}
\begin{equation}
	-\varepsilon_0 \frac{d^2\phi}{dr^2} + \frac{dp_r}{dr} = q(p_v - n_c + N_d^+ - N_a^-),
\end{equation}
\begin{equation}
	-\frac{dJ_n}{dr} + G - B(n_c p_v - N_i^2) = 0,
\end{equation}
\begin{equation}
	-\frac{dJ_p}{dr} + G - B(n_c p_v - N_i^2) = 0,
\end{equation}
\begin{equation}\label{Eqn: 1D Jn}
	J_n = - \frac{\nu_n k_B T}{q} \frac{dn_c}{dr} + \nu_n n_c \frac{d\phi}{dr},
\end{equation}
\begin{equation}\label{Eqn: 1D Jp}
	J_p = - \frac{\nu_p k_B T}{q} \frac{dp_v}{dr} - \nu_p p_v \frac{d\phi}{dr}, 
\end{equation}

where 

\begin{equation*}
	N_d^+ = N_d \bigg[ 1 - \frac{1}{1 + \frac{1}{2} \exp(\frac{-E_{F_m} + E_d - q\phi}{k_B T})} \bigg],
\end{equation*}
\begin{equation*} 
	N_a^- = N_a \bigg[ 1 + 2 \exp (\frac{-E_{F_m} + E_a - q \phi}{k_B T}) \bigg]^{-1},
\end{equation*}
\begin{equation*}
	N_i =\sqrt{N_c N_v} \exp ( -\frac{E_c - E_v}{2k_B T} ),
\end{equation*}

with boundary conditions

\begin{equation*}
	\frac{dp_r}{dr}(r=0) = \frac{dp_r}{dr}(r=L) = 0,
\end{equation*}
\begin{equation*}
	\frac{dp_s}{dr}(r=0) = \frac{dp_s}{dr}(r=L) = 0,
\end{equation*}
\begin{equation*}
	\frac{d\theta_r}{dr}(r=0) = \frac{d\theta_r}{dr}(r=L) = 0,
\end{equation*}
\begin{equation*}
	\frac{d\theta_s}{dr}(r=0) = \frac{d\theta_s}{dr}(r=L) = 0,
\end{equation*}
\begin{equation*}
	\phi(r = 0) = 0, \quad \phi(r = L) = 0,
\end{equation*}
\begin{equation*}
	n_c(r = 0) = n_c(r = L) = N_c e^{-(E_c - E_{f_m}) / k_B T},
\end{equation*}
\begin{equation*}
	p_v(r = 0) = p_v(r = L) = N_v e^{-(E_{f_m} - E_v) / k_B T}.
\end{equation*}

%-----------------------------------------------------------------------------------
\subsection{Numerical issues} 
%-----------------------------------------------------------------------------------
The model derived above comprises of differential equations that are nonlinear and coupled, which can prove troublesome numerically. So we non-dimensionalize the problem as in Appendix \ref{app:nondim}.  Further, we notice that the coupling between the first five governing equations and the rest of the model is weak.  This is especially so when the length of the simulated device is much smaller than the Debye length, or when the dimensionless quantity $\delta$ is small, which is generally the case in the simulations in this paper. Therefore we treat them as two subproblems, that are then solved self-consistently until convergence occurs. Each subproblem is constructed within the finite difference framework, and the resulting system of nonlinear equations is solved using the trust-region dogleg method.

%===================================================================================
\section{Application to Bismuth Ferrite}  \label{Section:App}
%===================================================================================

%-----------------------------------------------------------------------------------
\subsection{Material constants} \label{Appendix: BFO parameters}
%-----------------------------------------------------------------------------------

The coefficients of the Devonshire-Landau energy for BFO in equation (\ref{Eqn: GLD energy}) are presented in Table \ref{Table: LGD parameters}. They are derived to match the values of spontaneous polarization, tilt angles, and dielectric constant \cite{Dieguez2011,Dieguez2013,Kamba2007}.  Other material parameters including band structure information \cite{Clark2007} and carrier mobility values \cite{Scott2007} are listed in Table \ref{Table: parameters}. The values of $a_0$ and $b_0$ are chosen to match a ferroelectric domain wall width of 2 nm. Typically BFO exists as a n-type semiconductor due to oxygen vacancies. It can also become p-type with Bi deficiency. Here we restrict our simulations to n-type semiconductors.

\begin{table}
	\begin{center}
		\begin{tabular}{ |c|c|c| } 
			\hline
			Symbols & Values & Units \\ 
			\hline
			$a_{1}$		& $-1.19 \times 10^9$		& V$\,$m$\,$C$^{-1}$ \\
			$a_{11}$	& $9.93 \times 10^8$		& V$\,$m$^{5}\,$C$^{-3}$ \\
			$a_{12}$	& $3.93 \times 10^8$		& V$\,$m$^{5}\,$C$^{-3}$ \\
			$b_{1}$		& $-1.79 \times 10^{10}$	& V$\,$m$^{-3}\,$C \\
			$b_{11}$	& $1.14 \times 10^{11}$ 	& V$\,$m$^{-3}\,$C \\
			$b_{12}$	& $2.25 \times 10^{11}$		& V$\,$m$^{-3}\,$C \\
			$c_{11}$	& $1.50 \times 10^{10}$		& V$\,$m$\,$C$^{-1}$ \\
			$c_{12}$	& $7.50 \times 10^9$		& V$\,$m$\,$C$^{-1}$ \\
			$c_{44}$	& $-1.60 \times 10^{1-}$	& V$\,$m$\,$C$^{-1}$ \\
			\hline
		\end{tabular}
		\caption{Coefficients of Laudau-Devonshire energy for BFO}
		\label{Table: LGD parameters}
	\end{center}
\end{table}

\begin{table}
	\begin{center}
		\begin{tabular}{ |l|c|c|c| } 
			\hline
			Parameters & Symbols & Values & Units \\ 
			\hline
			Electron mobility 			& $\mu_n$ 	& $2 \times 10^{-5}$ 	& m$^2\,$V$^{-1}$ s$^{-1}$ \\ 
			Hole mobility 				& $\mu_p$ 	& $1 \times 10^{-5}$ 	& m$^2\,$V$^{-1}$ s$^{-1}$ \\ 
			Energy of conduction band 	& $E_c$ 	& $-3.3$ 				& eV \\ 
			Energy of valence band 		& $E_v$ 	& $-6.1$ 				& eV \\
			Donor level 				& $E_d$ 	& $-3.7$ 				& eV \\
			Acceptor level 				& $E_a$ 	& $-5.8$ 				& eV \\
			Effective density of states for conduction band& $N_c$ & $1 \times 10^{24}$ & m$^{-3}$ \\
			Effective density of states for valence band& $N_v$ & $1 \times 10^{24}$ & m$^{-3}$ \\
			Donor concentration 		& $N_d$ 	& $1 \times 10^{20}$ 	& m$^{-3}$ \\
			Acceptor concentration 		& $N_a$ 	& $0$ 					& m$^{-3}$ \\
			Polarization gradient coefficient & $a_0$ & $9 \times 10^{-10}$ & V$\,$m$^{3}$ C$^{-1}$ \\
			AFD gradient coefficient 	& $b_0$ 	& $2 \times 10^{-9}$ 	& V$\,$m$^{-1}\,$C \\
			Rate of photogeneration 	& $G$ 		& $1 \times 10^{27}$ 	& m$^{-3}$ s$^{-1}$ \\ 
			Radiative recombination coefficient	& $B$ & $1 \times 10^{-9}$	& m$^{3}$ s$^{-1}$ \\ 
			Thickness of film 			& $L$ 		& $100$				 	& nm \\ 
			Temperature 				& $T$ 		& $300$ 				& K \\ 
			Work function of Pt 		& $-E_{Fm}$ & $5.3$ 				& eV \\  
			\hline
		\end{tabular}
		\caption{Material and simulation parameters}
		\label{Table: parameters}
	\end{center}
\end{table}

%-----------------------------------------------------------------------------------
\subsection{Two-domain ferroelectrics}
%-----------------------------------------------------------------------------------
%-----------------------------------------------------------------------------------
\subsubsection{71$\degree$ and 109$\degree$ domain walls}
%-----------------------------------------------------------------------------------

\begin{table}
	\centering
	\begin{tabular}{ C{3.2cm}  C{3.2cm}  C{3.2cm}  C{3.2cm} }
		%\cline{2-9}
		\hline
		\addlinespace
		\multicolumn{2}{c}{\includegraphics[width=0.15\textwidth]{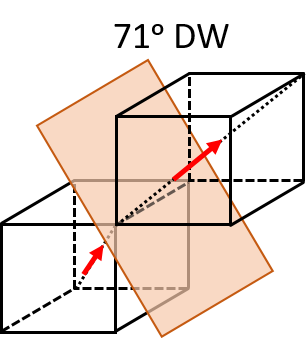}} 
		& \multicolumn{2}{c}{\includegraphics[width=0.15\textwidth]{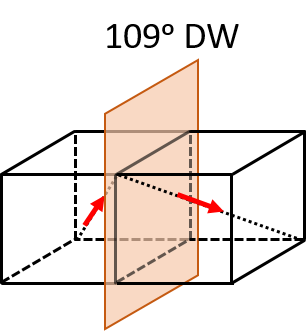}}\\
		\multicolumn{2}{c}{Polarization: $[1 1 1] \longrightarrow [1 1 \overline{1}]$} 
		& \multicolumn{2}{c}{Polarization: $[1 1 1] \longrightarrow [1 \overline{1} \overline{1}]$}  \\
		\addlinespace
		\hline
		\addlinespace
		\textbf{(a)} & \textbf{(b)} & \textbf{(c)} & \textbf{(d)} \\
		Continuous OT & Discontinuous OT & Continuous OT & Discontinuous OT \\
		$ \langle 1 1 1 \rangle \longrightarrow \langle 1 1 \overline{1} \rangle $ &
		$ \langle 1 1 1 \rangle \longrightarrow \langle \overline{1} \overline{1} 1 \rangle $ &
		$ \langle 1 1 1 \rangle \longrightarrow \langle 1 \overline{1} \overline{1} \rangle $ &
		$ \langle 1 1 1 \rangle \longrightarrow \langle \overline{1} 1 1 \rangle $ \\
		$E_{DW} = 0.53$ J$\,$m$^{-2}$ & $E_{DW} = 0.63$ J$\,$m$^{-2}$ & $E_{DW} = 0.53$ J$\,$m$^{-2}$ & $E_{DW} = 0.45$ J$\,$m$^{-2}$ \\
		$J_{sc} = -0.22$ A$\,$m$^{-2}$ & $J_{sc} = -0.84$ A$\,$m$^{-2}$ & $J_{sc} = 0.98$ A$\,$m$^{-2}$ & $J_{sc} = -1.0$ A$\,$m$^{-2}$ \\
		$V_{oc} = 6.8 \text{ mV}$ & $V_{oc} = 29 \text{ mV}$ & $V_{oc} = -70 \text{ mV}$ & $V_{oc} = 38 \text{ mV}$ \\
		\addlinespace[1mm]
		\hline
		\addlinespace
		\multicolumn{4}{l}{* $[\cdot] \longrightarrow [\cdot]$ and $\langle \cdot \rangle \longrightarrow \langle \cdot \rangle$ denote the directions of electric polarization and oxygen octahedra} \\
		\multicolumn{4}{l}{tilt (OT), respectively, on two neighboring domains. $J_{sc}$ and $V_{oc}$ are the short-circuit current} \\
		\multicolumn{4}{l}{density and open-circuit voltage obtained from our device model simulations. $E_{DW}$ refers to} \\
		\multicolumn{4}{l}{the domain wall energy calculated at thermal equilibrium in the absence of light illumination.} \\
	\end{tabular}
	\caption{Device models with different types of domain walls. }
	\label{Table: DW types}
\end{table}

We consider a device comprising of a BFO film with two ferroelectric domains separated by either a $71\degree$ or $109\degree$ domain wall, with continuous or discontinuous oxygen octahedra rotations across the DW. This gives a total of four different cases, as illustrated in Table \ref{Table: DW types}. 

\begin{figure}
	\begin{center}
		\includegraphics[width=0.7 \textwidth]{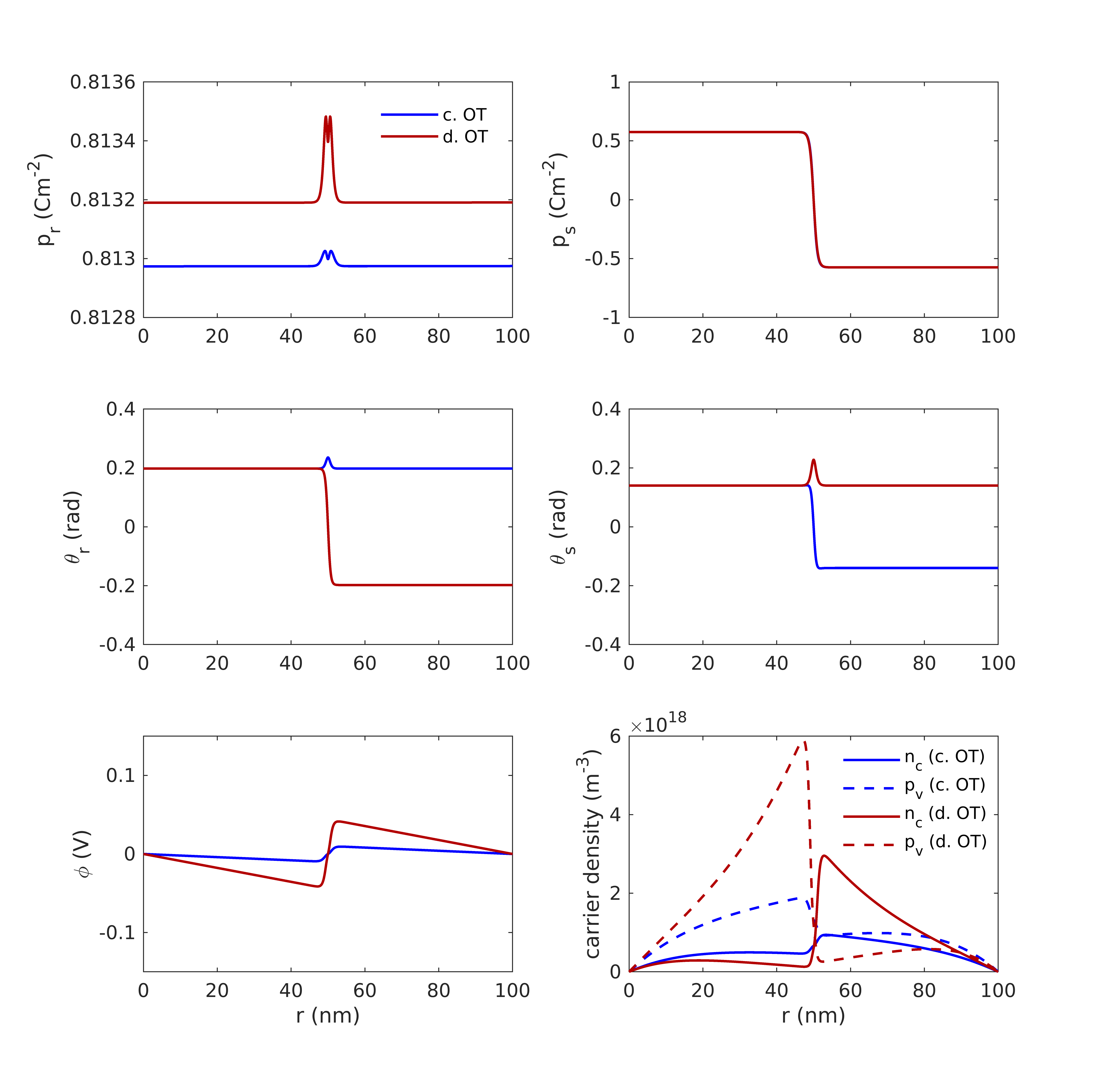}
	\end{center}
	\caption{Spatial variation of field quantities (polarization components, OT tilt angles, electric potential and carrier densities) along the $71\degree$ DW device with either continuous or discontinuous OT at short circuit under light illumination}
	\label{Fig: 71deg}
\end{figure}

\begin{figure}
	\begin{center}
		\includegraphics[width=0.7 \textwidth]{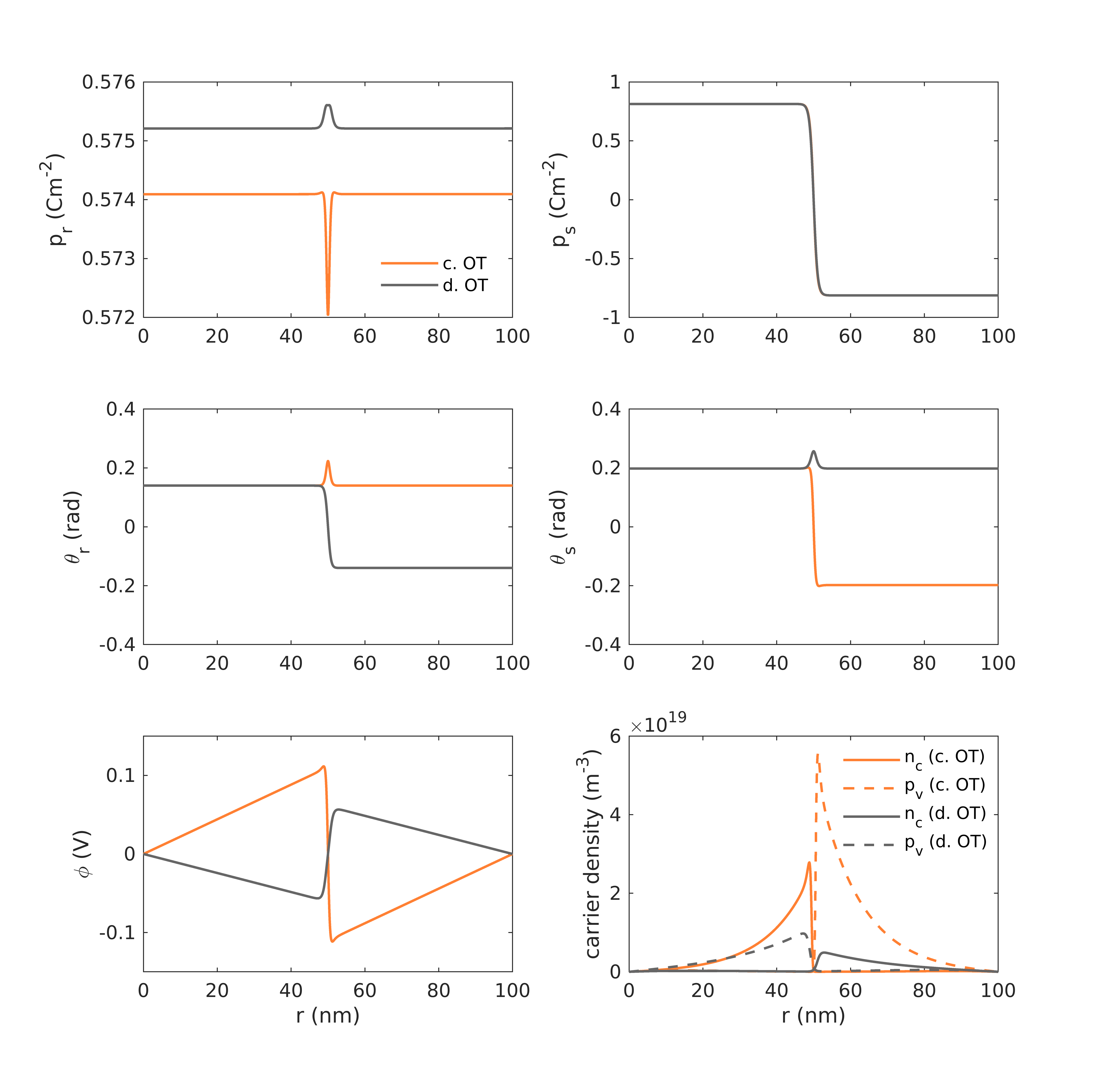}
	\end{center}
	\caption{Spatial variation of field quantities (polarization components, OT tilt angles, electric potential and carrier densities) along the $109\degree$ DW device with either continuous or discontinuous OT at short circuit under light illumination}
	\label{Fig: 109deg}
\end{figure}

\begin{figure}
	\begin{center}
		\includegraphics[width=0.6 \textwidth]{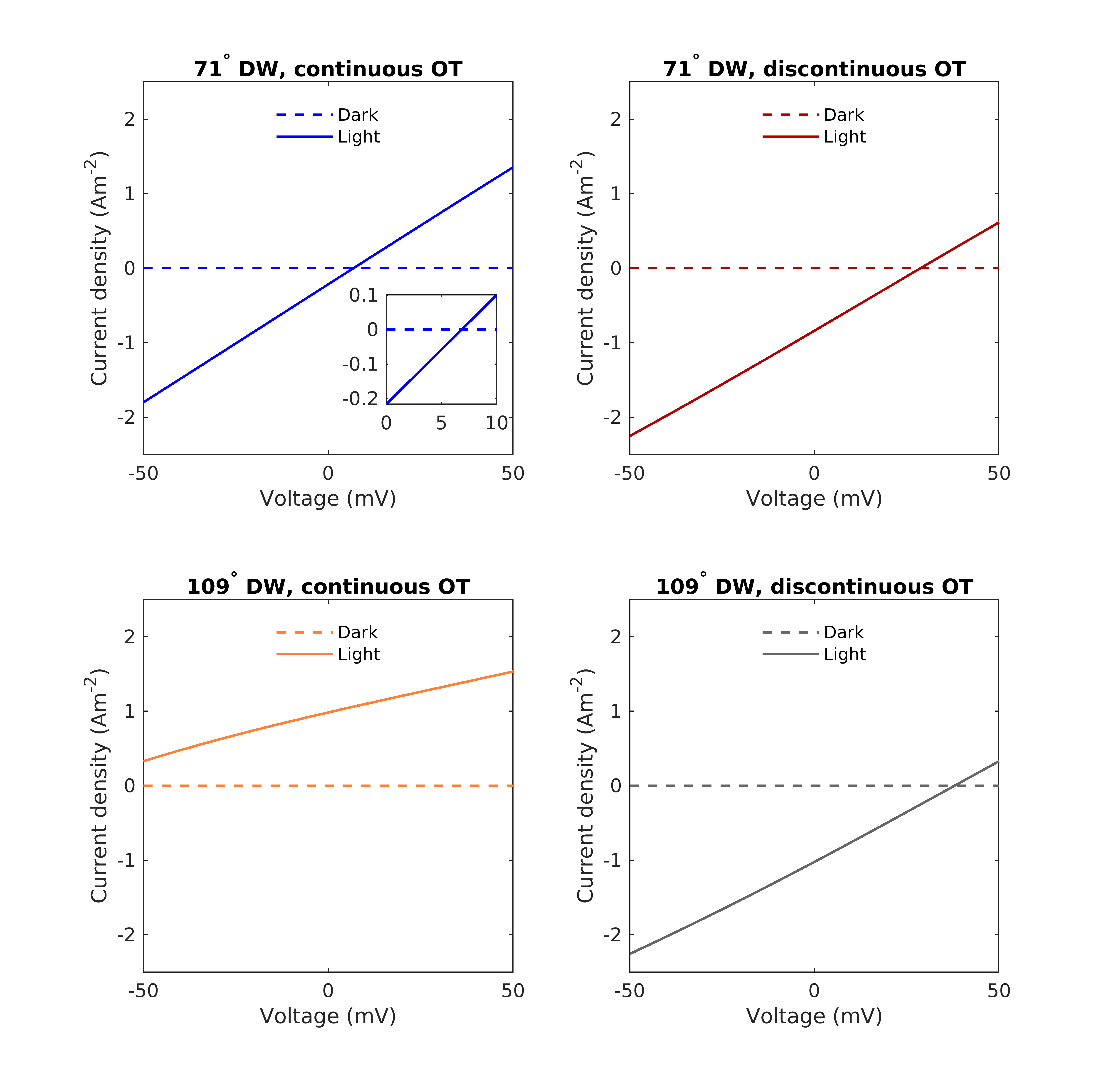}
	\end{center}
	\caption{Current-voltage plot in dark ($G = 0$) and under light illumination ($G = 10^{27}\,$m$^{-3}\,$s$^{-1}$) for BFO devices with a $71 \degree$ or $109 \degree$ DW with continuous or discontinuous OT.}
	\label{Fig: IV plot}
\end{figure}

Figures \ref{Fig: 71deg} and \ref{Fig: 109deg} show the variation of various field quantities when the perovskite film is exposed to light illumination and shorted.  Notice that in all cases, the perpendicular component of the polarization $p_r$ is not constant in the vicinity of the domain wall.  In other words, the polarization is not divergence free, and we see a voltage drop across the domain wall.  The polarization profile  (i.e. $p_r$) of $71 \degree$ and $109 \degree$ domain walls with continuous OT are qualitatively similar to those obtained from first-principles calculations \cite{Lubk2009}.  This voltage drop across the domain wall leads to charge separation of photogenerated electron-hole pairs, and a non-zero photocurrent.  This is evident in the current-voltage plots shown in Figure \ref{Fig: IV plot} and is consistent with the mechanism proposed by  Yang \textit{et al.} \cite{Yang2010}.

Figure \ref{Fig: IV plot} shows that the magnitude and direction of photocurrent due to the domain wall effect hinge greatly upon the changes in the crystallographic structure across the domain wall.   The case with $109 \degree$ DW and continuous OT gives a positive short-circuit current, which is in the same direction as net polarization in the film while the rest show negative currents.  Importantly, this DW-based photovoltaic effect shows that the direction of current flow does not necessarily correlates with the direction of net polarization consistent with experimental observations \cite{Inoue2015}.

Furthermore, we observe strong coupling between polarization and oxygen octahedra tilt (OT). This is evident from the vastly different results (including the change in current direction) obtained when changing the OT profiles without changing the type of domain walls. In actual experiments, we may only observe one type of oxygen octahedra rotation for each domain wall type. We compute the domain wall energy for each case as in Table \ref{Table: DW types}, and find that it is energetically more favorable to have $71 \degree$ domain wall with continuous OT and $109 \degree$ domain wall with discontinous OT. This is consistent with previous first-principles calculations \cite{Dieguez2013, Xue2014}.

\begin{figure}
	\begin{center}
		\includegraphics[width=0.4 \textwidth]{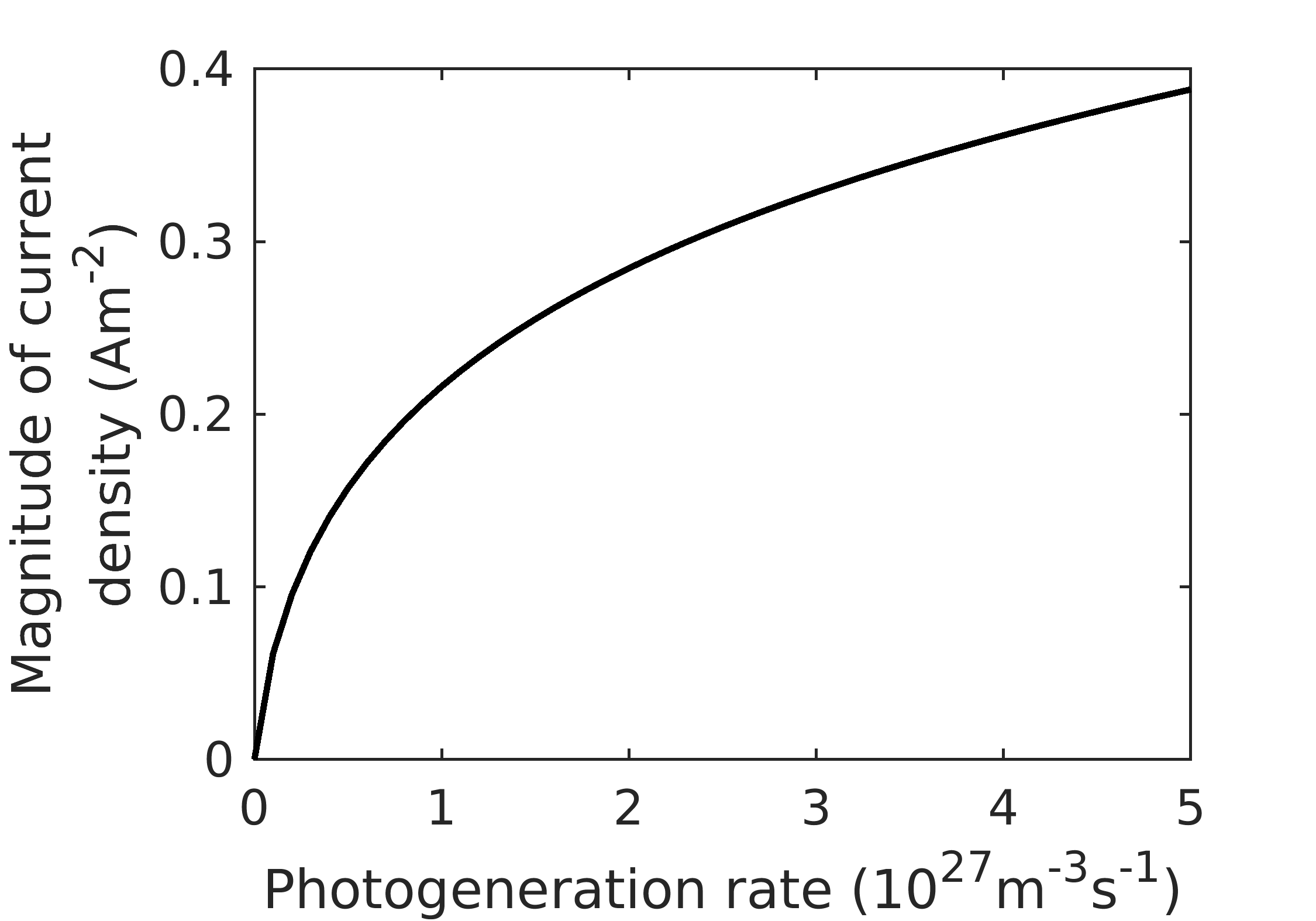}
	\end{center}
	\caption{Short-circuit current density versus photogeneration rate for a two-domain ferroelectrics separated by a 71$\degree$ DW with continuous OT.}
	\label{Fig:IG}
\end{figure}

As the perovskite film is first exposed to light illumination, which is simulated in terms of an increase in photogeneration rate of electron-hole pairs, the magnitude of short-circuit current density generated increases rapidly initially as shown in Figure \ref{Fig:IG}. The increase slows down at higher illumination (or photogeneration rate) due to recombination of the excited electrons and holes.

\begin{figure}
	\begin{center}
		\includegraphics[width=0.7 \textwidth]{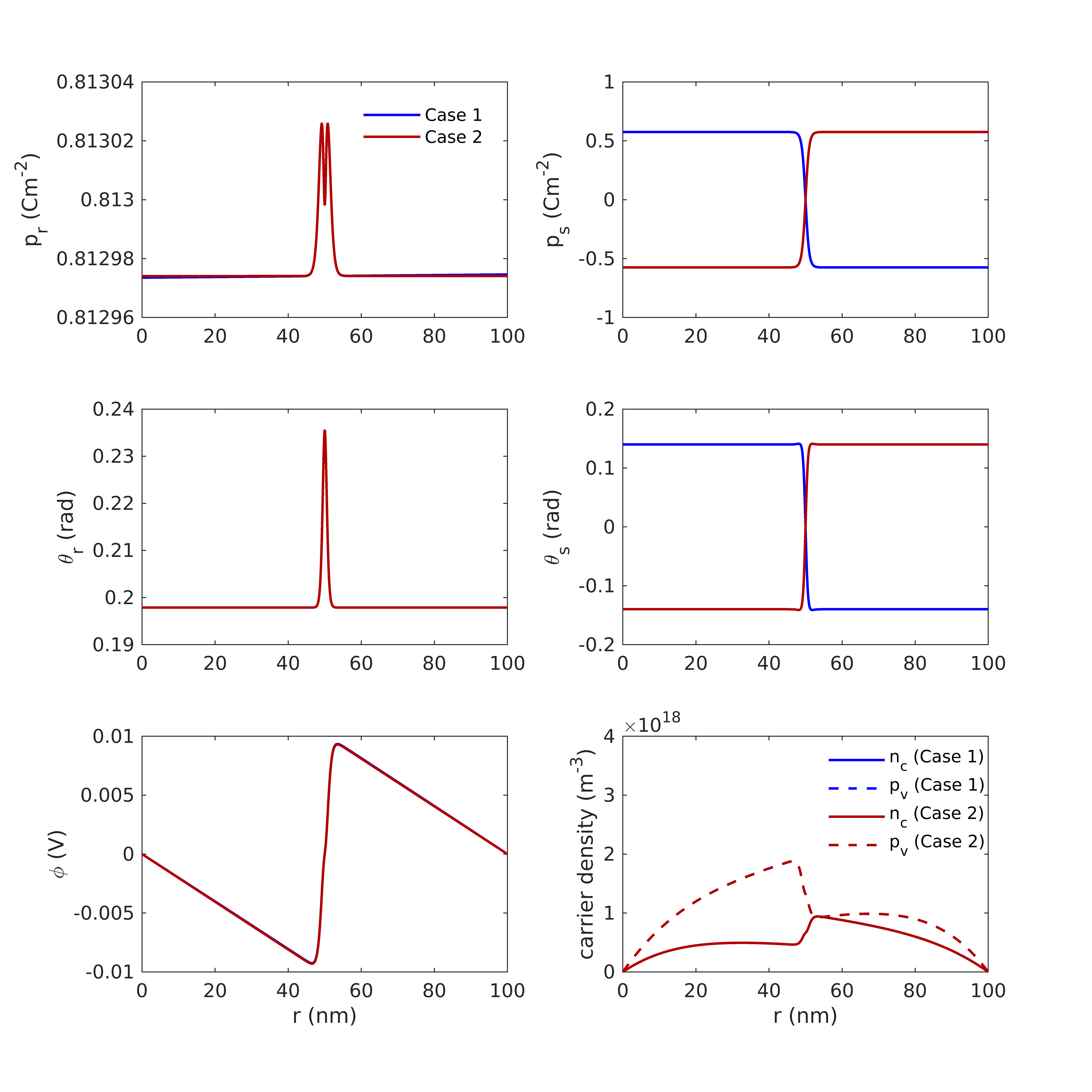}
	\end{center}
	\caption{Spatial variation of field quantities along a two-domain device continuous ($71 \degree$ DW) at short circuit. The two cases are identical except the order of the two domains is reversed.}
	\label{Fig: Swapped domains}
\end{figure}

Finally we consider changing the order of the domains in a two-domain device. Figure \ref{Fig: Swapped domains} shows that doing so does not pose any difference to the profiles of other quantities such as $p_r$ and $\phi$. This implies that current flows in a single direction irrespective of the order of the domains. If we were to stack the different domains to form a device with periodic domain pattern (i.e. alternating domains), the photovolatic effect would be additive and would not cancel out. This is exactly what we observe in Section \ref{Section: multiple DWs}.

%-----------------------------------------------------------------------------------
\subsubsection{180$\degree$ domain walls}
%-----------------------------------------------------------------------------------
In the case of the 180$\degree$ domain walls, with either continuous or discontinuous OTs, there is no visible disturbance to the polarization component normal to the domain wall at the domain wall. With a lack of symmetry breaking, the photovoltaic effect fails to be generated. The figures are omitted for brevity.

%-----------------------------------------------------------------------------------
\subsection{Ferroelectrics with multiple domain walls} \label{Section: multiple DWs}
%-----------------------------------------------------------------------------------

\begin{figure}
	\begin{center}
		\includegraphics[width=0.7 \textwidth]{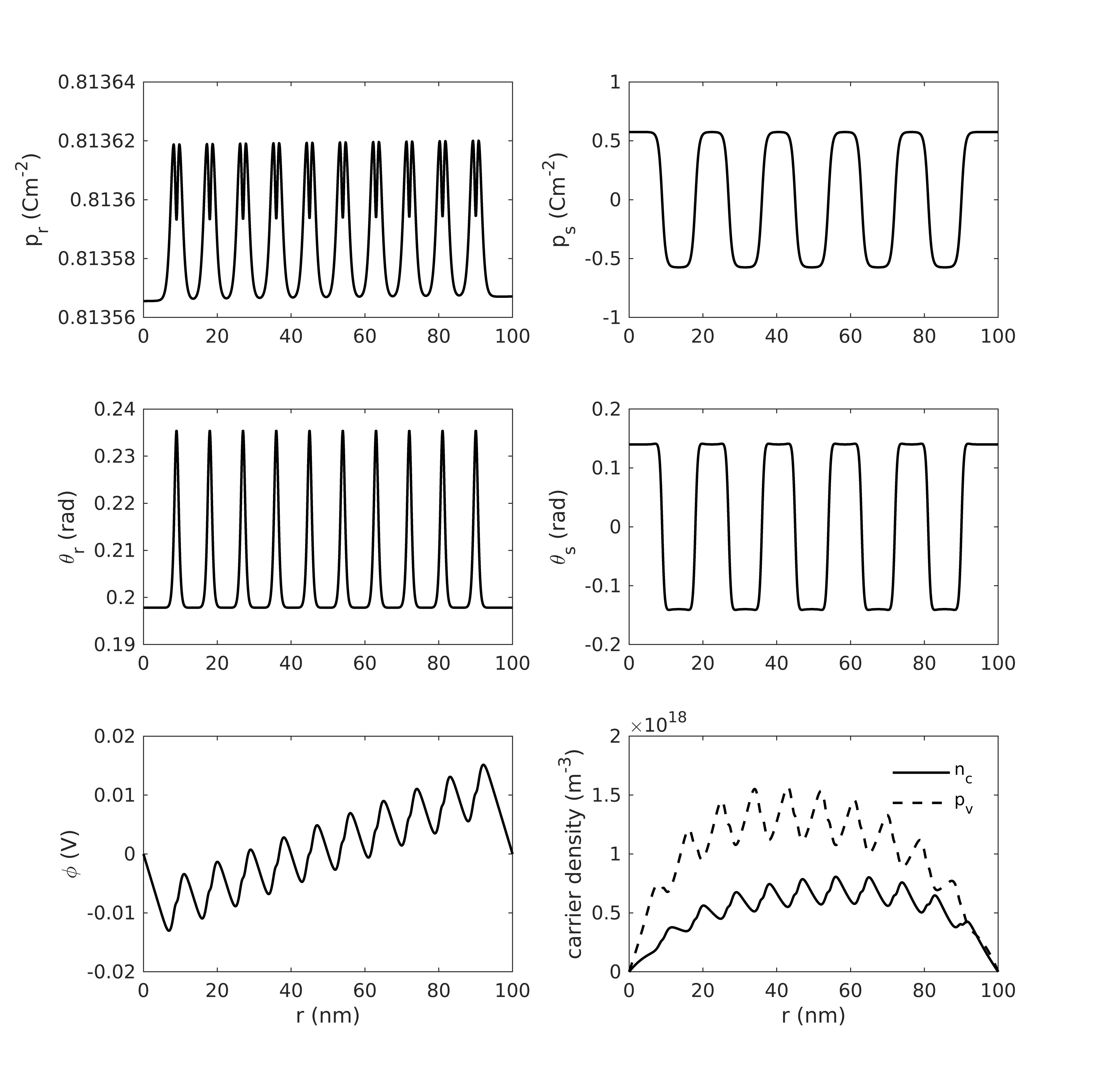}
	\end{center}
	\caption{Spatial variation of field quantities for ferroelectrics with ten 71$\degree$ DWs and continuous OTs at short circuit.}
	\label{Fig: 10 DWs}
\end{figure}

We now examine the case with multiple domain walls. We keep the width of the perovskite film constant and uniformly place a number of domain walls parallel to the metal electrodes within the film. The distribution of polarization, oxygen octahedra tilts (OTs) and other field quantities for a shorted device with ten 71$\degree$ domain walls are presented in Figure \ref{Fig: 10 DWs}. Polarization and OTs are periodic and electric potential varies in a zig-zag manner but with a slope. 

\begin{figure}
	\begin{center}
		\includegraphics[width=0.4 \textwidth]{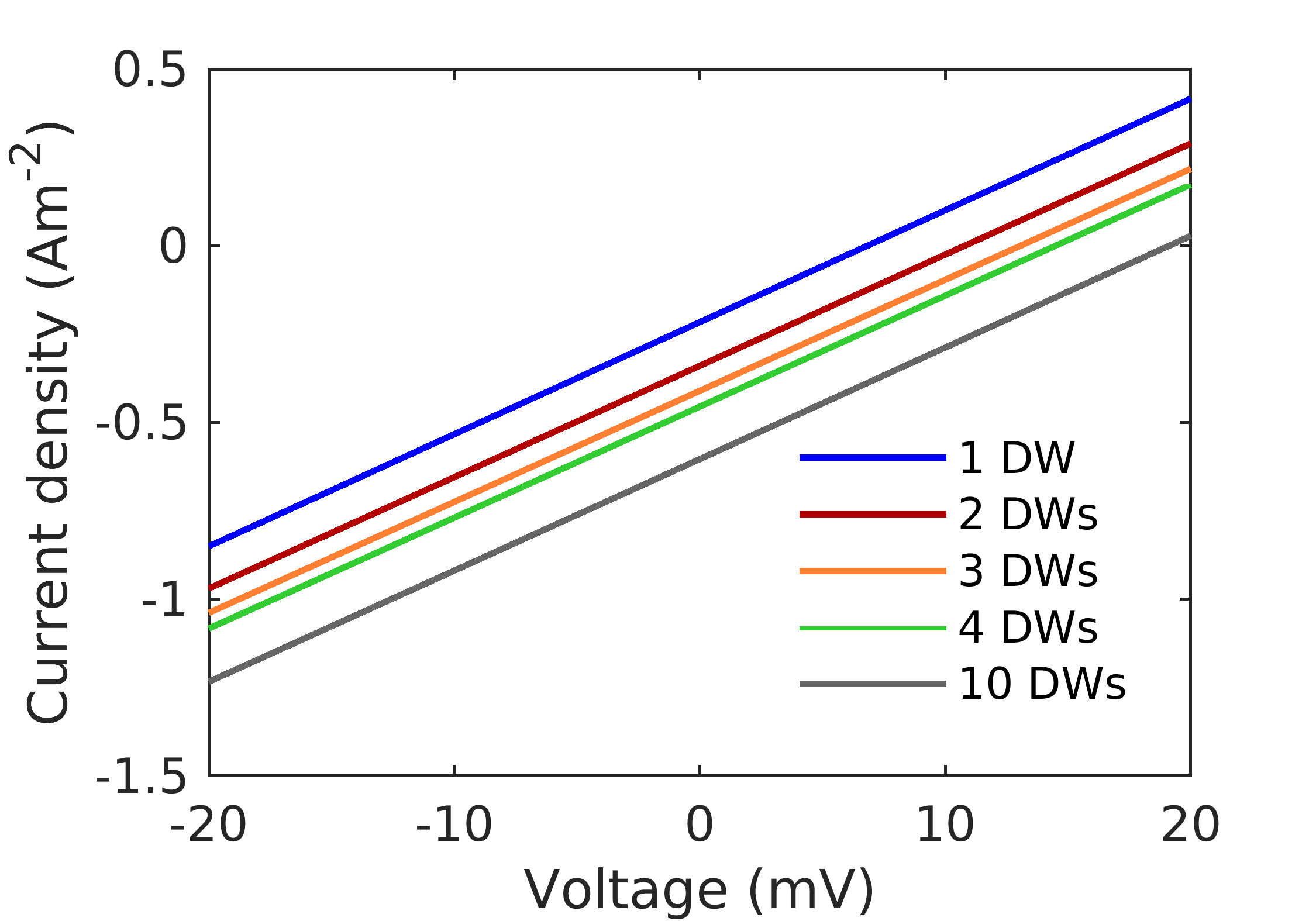}
	\end{center}
	\caption{Current-voltage plot for multi-domain BFO devices with $71\degree$ DWs and continuous OTs}
	\label{Fig: Multi-domain}
\end{figure}

Figure \ref{Fig: Multi-domain} shows that the magnitudes of both short-circuit current and open-circuit voltage increase with the density of domain walls in the device. The additive effect becomes smaller at higher domain wall number as it is influenced by the boundary.

\newpage
%-----------------------------------------------------------------------------------
\subsection{Effect of varying doping and width of ferroelectric film}
%-----------------------------------------------------------------------------------

\begin{figure}
	\begin{center}
		\subfigure[Perovskite width of 100 nm]{
			\includegraphics[width=0.6 \textwidth]{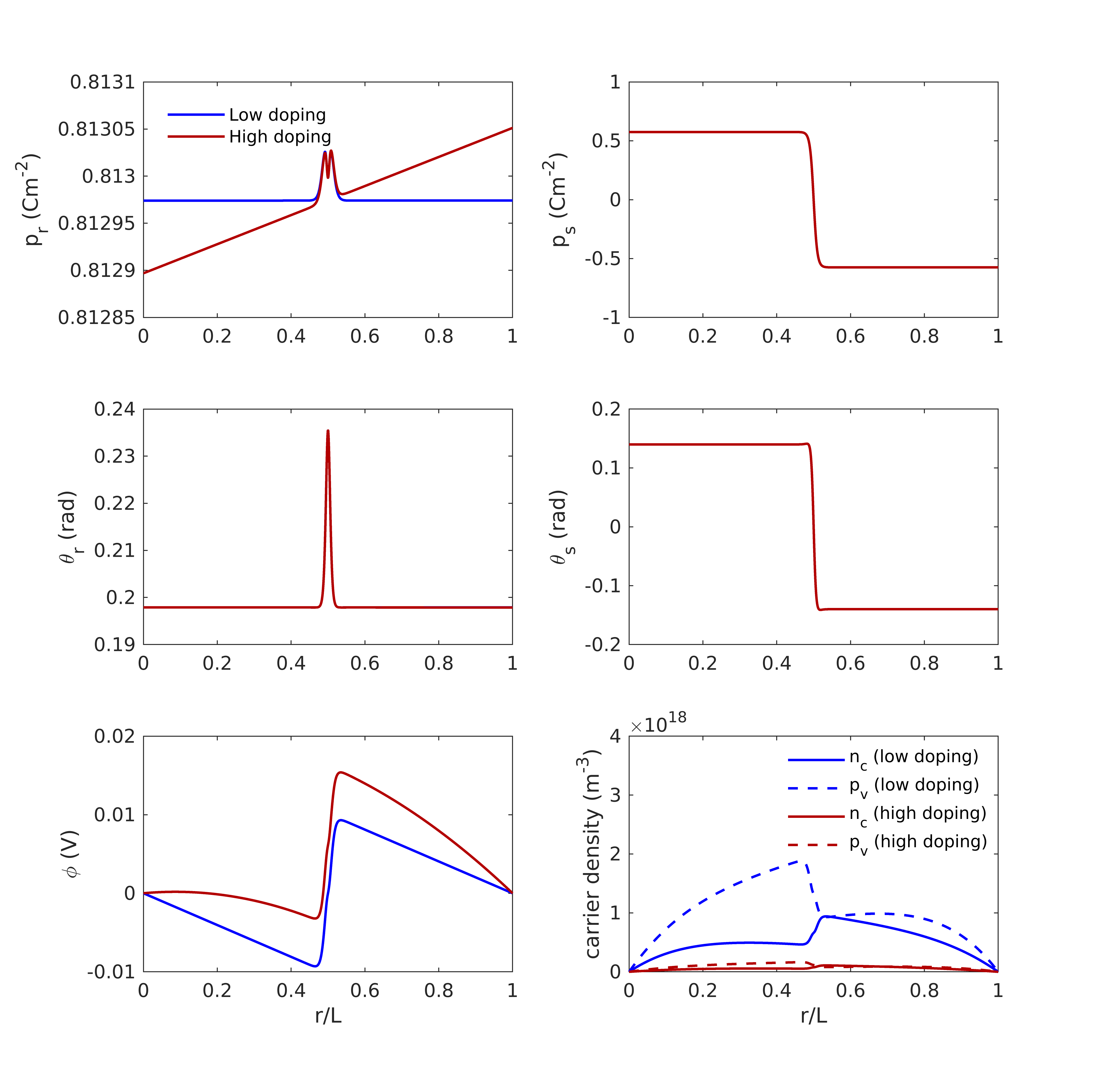}
		}
		\subfigure[Perovskite width of 500 nm]{
			\includegraphics[width=0.6 \textwidth]{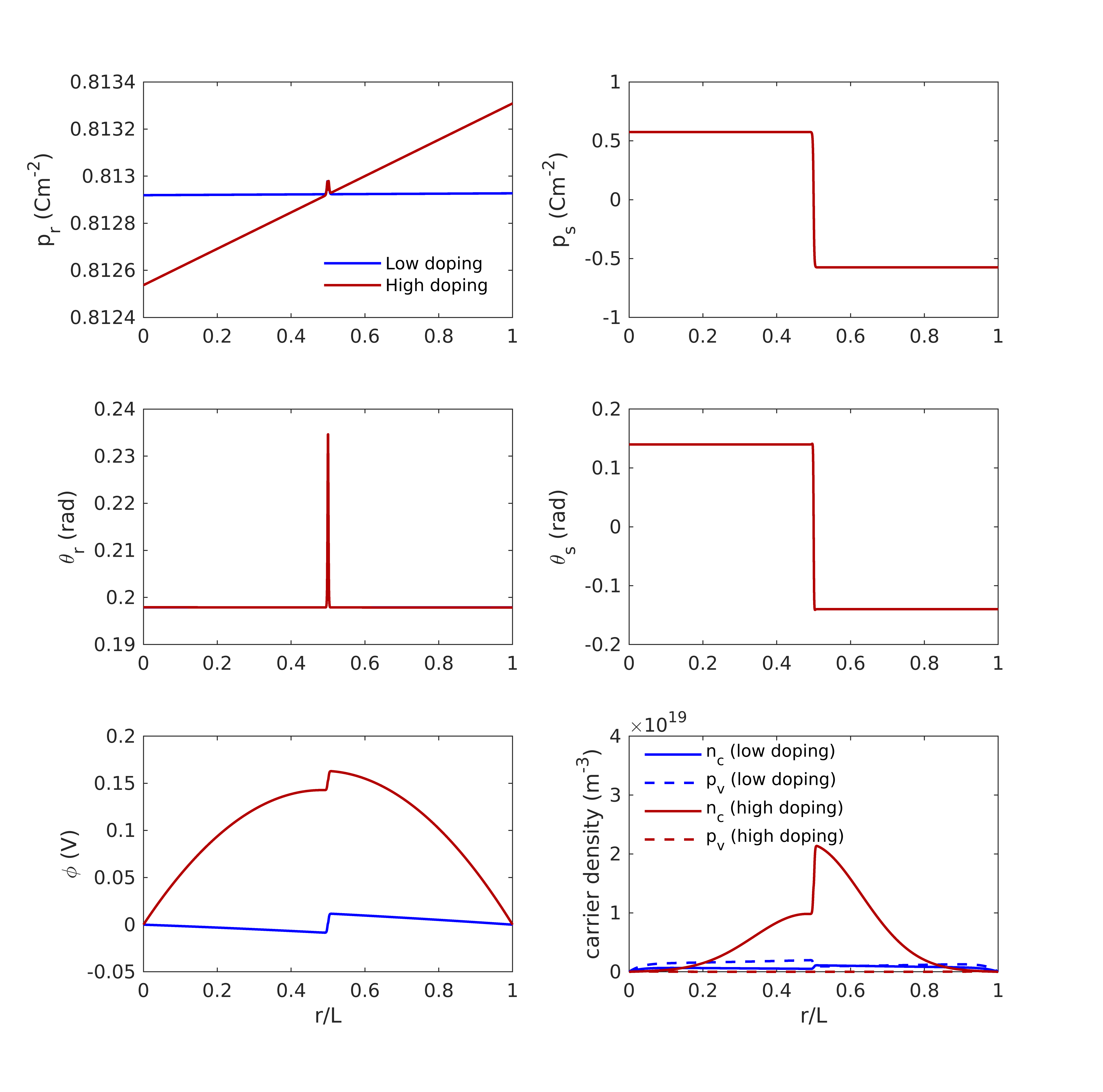}
		}
	\end{center}
	\caption{Low dopant level: $N_d = 10^{20} \, \text{m}^{-3}$, high dopant level: $N_d = 10^{22} \, \text{m}^{-3}$}
	\label{Fig: Doping and width}
\end{figure}

Next, we investigate the effect of doping and width on the ferroelectric response using a two-domain example. All the previous simulations are run using a small donor doping density of $N_d = 10^{20} \, \text{m}^{-3}$ and a width of 100 nm, which corresponds to the state of complete depletion. Typically a depletion layer forms at a metal-semiconductor interface and the width of the depletion layer is related to the Debye length which is dependent on the dopant density and dielectric constant. Complete depletion occurs when the Debye length of the material is much larger than the width of the device. Otherwise there is partial or local depletion. Figure \ref{Fig: Doping and width} shows the short-circuit distributions for two different perovskite widths of 100 nm and 500 nm at a low dopant density level of $N_d = 10^{20} \, \text{m}^{-3}$ and a high dopant density of $N_d = 10^{22} \, \text{m}^{-3}$. At a small width of 100 nm, changing the doping level from low to high does minimal changes to the field profiles and the photovolatic response, with its short-circuit current density staying almost constant at -0.22$\,$A$\,$m$^{-2}$. On the other hand, at a larger width of $500\,$nm, increasing the doping level changes the state of the perovskite film from complete depletion to partial depletion, and at the same time, raises the short-circuit current density from -0.011$\,$A$\,$m$^{-2}$ to -0.033$\,$A$\,$m$^{-2}$. The resulting electric potential profile can be viewed as the superposition of two contributions: domain walls and depletion. This illustrates the feature of the model in combining the ferroelectric and semiconductor behavior of the material.

%===================================================================================
\section{Conclusions and Discussion} \label{Section:Conc}
%===================================================================================

In this paper, we have proposed a thermodynamically consistent continuum device model to study photovoltaic effect in multi-domain ferroelectric perovskites accounting systematically for the interactions among space charge, polarization and oxygen octahedra tilts. The model has been successfully implemented numerically. Our results show that there is an electric potential step across each 71$\degree$ or 109$\degree$ domain wall, and that this produces a PV effect.  There is no electric potential step across a 180$\degree$ domain wall, and correspondingly no PV effect.  Further, the model shows that the direction of current depends on the nature of the domain wall and not the orientation of domains.  Therefore,  the PV effect becomes additive across multiple domain walls with alternating domains.

\begin{figure}
	\begin{center}
		\includegraphics[width=0.48 \textwidth]{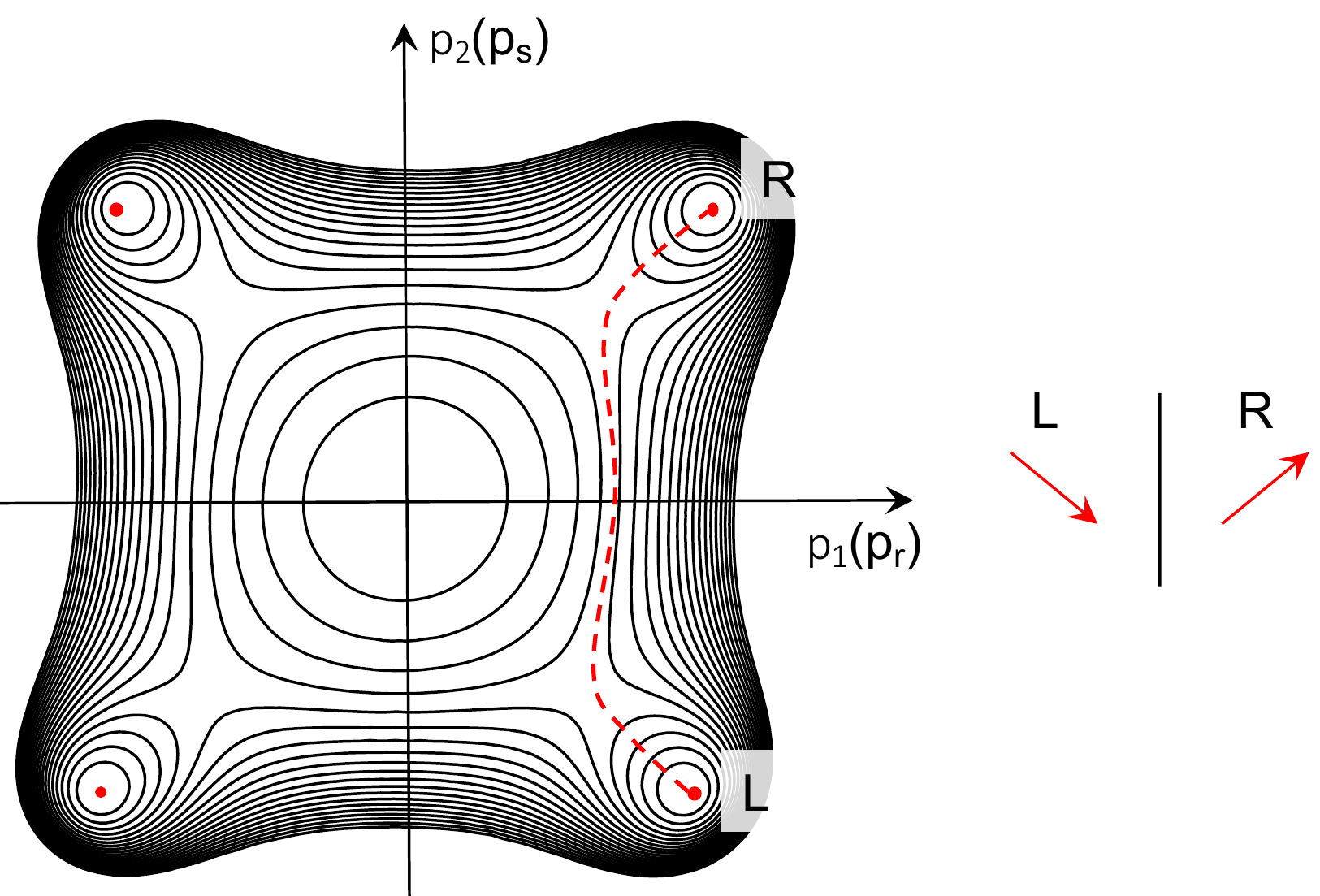}
	\end{center}
	\caption{The energy landscape of a non-180$\degree$ domain wall.}
	\label{Fig:generic}
\end{figure}

We note that the presence of electric potential step across non-180$\degree$ or the lack of such a step for a 180$\degree$ domain wall is a generic feature.  Consider a generic Devonshire-Landau energy landscape shown in Figure \ref{Fig:generic}. Further, consider a non-180$\degree$ domain wall that separates two ferroelectric domains, one with polarization $L$ and the other with polarization $R$, as marked in the same figure. These two polarizations vectors are spontaneous polarizations, and are thus energy minima of the Devonshire-Landau energy.  Now, as the polarization changes from the value $L$ to the value $R$ or vice versa across the domain wall, it will do so along the low energy valley as shown by the dashed line. This path necessarily involves a change in the component of polarization normal to the domain wall.   Therefore, there will indeed be a electric potential step across this domain wall.  Note that the electric potential step depends on the path connecting the two polarizations, and this is unchanged if the domains are swapped.  This argument is generic because there is no reason in symmetry for the low energy valley to go in a straight line from $L$ to $R$.  A similar argument shows the lack of such a step for a 180$\degree$ domain wall is also a generic feature.  While the presence or absence of the step is a generic feature, its magnitude and direction depend on the specific energy landscape. 

%We note that the presence of electric potential step across non-180$\degree$ or the lack of such a step for a 180$\degree$ domain wall is a generic feature.  Consider a generic Devonshire-Landau energy landscape shown in Figure \ref{Fig:generic}, and consider the non-180$\degree$ shown on the right.  The two polarizations vectors (marked $L$ and $R$)  are spontaneous polarizations, and are thus energy minima of the Devonshire-Landau energy.  Further, the domain wall separating them necessarily has overall orientation as shown.  Now,  the polarization has to change from the value $L$ to the value $R$ marked in the figure as it goes across the domain wall.   It will do so along the low energy valley as shown by the dashed line.   Note that this path necessarily involves a change in the component of polarization normal to the domain wall.   Therefore, there will indeed be a electric potential step across this domain wall.  Note further that the electric potential step depends on the path connecting the two polarizations, and this is unchanged if the domains were swapped.  Finally, note that this argument is generic because there is no reason in symmetry for the low energy valley to go in a straight line from $L$ to $R$.  A similar argument shows the lack of such a step for a 180$\degree$ domain wall is also a generic feature.  While the presence or absence of the step is a generic feature, the magnitude and details depends on the specific energy landscape.  

The model and results presented here support the hypothesis that non-180$\degree$ domain walls contribute to the photovoltaic effect.  Importantly, this model does not {\it a priori} assume the domain wall structure or the electrostatic potential step across it.  Instead, this is a prediction of the model that is based on well-established Devonshire-Landau models of ferroelectrics and lumped band models of semiconductors.  This model is agnostic about the bulk photovoltaic effect.  We could modify our model to include it by coupling the photogeneration to the polarization, but we chose not to do so since this would not be predictive.

We note that it remains an open question as to why Alexe and Hesse  \cite{Alexe2011} do not see any difference in photocurrent in their AFM-based measurement.  One possibility is that the AFM-tip created a depletion zone around it which dominated over the potential step across the domain wall.  This can be investigated further by the proposed model but it requires a multi-dimensional numerical implementation that is the topic of current work.  

Finally, we  hope that the model presented here as well as the multi-dimensional numerical implementation of it will prove useful for future device designs.

%===================================================================================
% Acknowledgement
%===================================================================================
\section*{Acknowledgement}
Ying Shi Teh gratefully acknowledges the support of the Resnick Institute at Caltech through the Resnick Graduate Research Fellowship.

%===================================================================================
% Appendices
%===================================================================================
\begin{appendices}
	
%-----------------------------------------------------------------------------------
\section{Derivation of a thermodynamically consistent theory} \label{Appendix: dissipation rate}
%-----------------------------------------------------------------------------------

We outline a derivation of the model in Section \ref{Section:Theory}, and show that it is thermodynamically consistent.  Since we consider only isothermal processes, the second law of thermodynamics (Clausius-Duhem inequality) requires that the rate of dissipation be non-negative.  This rate of dissipation is given by 
\begin{equation}
\mathcal{D} = \mathcal{F} - \frac{d\mathcal{E}}{dt}
\end{equation}
where $\mathcal{F}$ is the rate of external work done on the system
\begin{equation}\label{Eqn: External work}
	\begin{split}
		\mathcal{F} = 
		\int_{\Omega} \mu_n G_{\texttt{net}} dV 
		+ \int_{\Omega} \mu_p G_{\texttt{net}} dV
		+ \frac{d}{dt} \int_{\partial\Omega_1 \cup \partial\Omega_2} \phi \sigma \, dS 
		- \int_{\partial\Omega} \mu_n \mathbf{J}_n \cdot \mathbf{\hat{n}} \, dS
		- \int_{\partial\Omega} \mu_p \mathbf{J}_p \cdot \mathbf{\hat{n}} \, dS.
	\end{split}
\end{equation}
and $\mathcal{E}$ is the energy stored in the system
\begin{equation}\label{Eqn: E}
	\mathcal{E} = \int_{\Omega} \left( W + \frac{\varepsilon_0}{2} |{\nabla \phi}|^2 \right) dV,
\end{equation}
The first two terms in Equation (\ref{Eqn: External work}) denote the rate of work done by incident photons in generating electron-hole pairs. Here $G_{\texttt{net}} = G - R$ denotes the net rate of photogeneration. The third term refers to the work done by the external voltage and $\sigma = \llbracket -\varepsilon_0 \nabla \phi + \chi \mathbf{p} \rrbracket \cdot \mathbf{\hat{n}}$ is the surface charge density where $\llbracket \cdot \rrbracket$ indicates a jump in the respective quantity $(\cdot)$ and $\mathbf{\hat{n}}$ is a unit vector normal to the surface.  The final two terms are the energy carried into the systems by electron and hole fluxes at the boundary.  The total energy consists of the free energy and the electrostatic energy stored in the electrostatic field.

Applying the divergence theorem and transport equations (\ref{Eqn: Transport nc}), (\ref{Eqn: Transport pv}), we rewrite equation (\ref{Eqn: External work}) as 

\begin{equation}\label{Eqn: External work 2}
\begin{split}
\mathcal{F} = 
\frac{d}{dt} \int_{\partial\Omega_1 \cup \partial\Omega_2} \phi \sigma dS 
- \int_{\Omega} \nabla \mu_n \cdot \mathbf{J}_n \, dV
- \int_{\Omega} \nabla \mu_p \cdot \mathbf{J}_p \, dV
- \int_{\Omega} \nabla \mu_{N_d^+} \cdot \mathbf{J}_{N_d^+} \, dV
- \int_{\Omega} \nabla \mu_{N_a^-} \cdot \mathbf{J}_{N_a^-} \, dV \\
+ \int_{\Omega} \mu_n \dot{n_c} \, dV
+ \int_{\Omega} \mu_p \dot{p_v} \, dV
+ \int_{\Omega} \mu_{N_d^+} \dot{N_d^+} \, dV
+ \int_{\Omega} \mu_{N_a^-} \dot{N_a^-} \, dV.
\end{split}
\end{equation}
From equations (\ref{Eqn: E}) and (\ref{Eqn: W}), the rate of change of the total energy of the system can be expressed as
\begin{equation}
\begin{split}
\frac{d\mathcal{E}}{dt}
= \int_{\Omega} \bigg( -\nabla \cdot \frac{\partial W_G}{\partial \nabla \mathbf{p}} + \frac{\partial W_{DL}}{\partial \mathbf{p}} \bigg) \cdot \dot{\mathbf{p}} \, dV
+ \int_{\Omega} \bigg( -\nabla \cdot \frac{\partial W_G}{\partial \nabla \mathbf{\theta}} + \frac{W_{DL}}{\partial \mathbf{\theta}} \bigg) \cdot \dot{\mathbf{\theta}} \, dV
+ \frac{d}{dt} \bigg( \frac{\varepsilon_0}{2} \int_{\mathbb{R}_3} |{\nabla \phi}|^2 dV \bigg) \\
+ \int_{\Omega} \bigg( \frac{\partial W_{p_v}}{\partial p_v} \dot{p_v}
+ \frac{\partial W_{n_c}}{\partial n_c} \dot{n_c}
+ \frac{\partial W_{N_d}}{\partial N_d^+} \dot{N_d^+}
+ \frac{\partial W_{N_a}}{\partial N_a^-} \dot{N_a^-} \bigg)\, dV \\
+ \int_{\partial \Omega} \bigg( \mathbf{\hat{n}} \cdot \frac{\partial W_G}{\partial \nabla \mathbf{p}} \bigg) \cdot \delta \mathbf{p} \, dS
+ \int_{\partial \Omega}  \bigg( \mathbf{\hat{n}} \cdot \frac{\partial W_G}{\partial \nabla \mathbf{\theta}} \bigg) \cdot \delta \mathbf{\theta} \, dS.
\end{split}
\end{equation}
Following \cite{Suryanarayana2012}, we can show
\begin{equation}
\frac{d}{dt} \bigg( \frac{\varepsilon_0}{2} \int_{\Omega} |{\nabla \phi}|^2 dV \bigg)
= \int_{\partial\Omega_1 \cup \partial\Omega_2} \phi \, \dot{\sigma} \, dS
+ \int_{\Omega} \nabla \phi \cdot \dot{\mathbf{p}} \, dV
+ \int_\Omega \phi \,q \,(z_d \dot{N_d^+} - z_a \dot{N_a^-} - \dot{n_c} + \dot{p_v}) dV.
\end{equation}
Therefore,
\begin{equation}
\begin{split}
\mathcal{D} &= \int_{\Omega} \Bigg[ \bigg( \nabla \cdot \frac{\partial W_g}{\partial \nabla \mathbf{p}} - \frac{\partial W_{GLD}}{\partial \mathbf{p}} - \nabla \phi \bigg) \cdot \dot{\mathbf{p}}
+ \bigg( \nabla \cdot \frac{\partial W_g}{\partial \nabla \mathbf{\theta}} - \frac{W_{GLD}}{\partial \mathbf{\theta}} \bigg) \cdot \dot{\mathbf{\theta}} 
+ \bigg( \mu_n -\frac{\partial W_{n_c}}{\partial n_c} + q\phi \bigg) \dot{n_c} \\
& \qquad + \bigg(\mu_p -\frac{\partial W_{p_v}}{\partial p_v} - q\phi \bigg) \dot{p_v}
+ \bigg( \mu_{N_d^+} - \frac{\partial W_{N_d}}{\partial N_d^+} - qz_d\phi \bigg) \dot{N_d^+} 
+ \bigg( \mu_{N_a^-} - \frac{\partial W_{N_a}}{\partial N_a^-} + qz_a\phi \bigg) \dot{N_a^-} \\
& \qquad - \nabla \mu_n \cdot \mathbf{J}_n - \nabla \mu_p \cdot \mathbf{J}_p
- \nabla \mu_{N_d^+} \cdot \mathbf{J}_{N_d^+} - \nabla \mu_{N_a^-} \cdot \mathbf{J}_{N_a^-} \Bigg] dV \\
& \qquad + \int_{\partial \Omega} \Bigg[ \bigg( \mathbf{\hat{n}} \cdot \frac{\partial W_g}{\partial \nabla \mathbf{p}} \bigg) \cdot \dot{\mathbf{p}} + \bigg( \mathbf{\hat{n}} \cdot \frac{\partial W_g}{\partial \nabla \mathbf{\theta}} \bigg) \cdot \dot{\mathbf{\theta}} \Bigg] dS.
\end{split}
\end{equation}
Each of the above terms is a non-negative product of a generalized force and a generalized velocity or rate.   Assuming either over-damped dynamics or equilibrium gives rise to the equations in Section \ref{Section:Theory}.	

%-----------------------------------------------------------------------------------
\section{Non-dimensionalization and scaling} \label{app:nondim}
%-----------------------------------------------------------------------------------

Before solving for the model numerically, we introduce dimensionless variables through appropriate scalings as follows:

$$ p_r = p_0 \op_r $$
$$ p_s = p_0 \op_s $$
$$ r = L_0 \ovr $$
$$ W_{DL} = W_0 \oW_{DL} $$
$$ \phi = \phi_0 \ophi $$
$$ n_c = N_0 \on_c $$
$$ p_v = N_0 \op_v $$
$$ N_d^+ = N_0 \oN_d^+ $$
$$ N_a^- = N_0 \oN_a^- $$
$$ N_i = N_0 \oN_i $$
$$ J_n = J_0 \oJ_n $$
$$ J_p = J_0 \oJ_p $$
$$ J_{total} = q J_0 \oJ_{total} $$
$$ G = G_0 \oG $$,

where

$$ \{p_0, \theta_0\} = \argmin_{p,\theta} (W_{DL} |_{|p_1| = |p_2| = |p_3| = p, |\theta_1| = |\theta_2| = |\theta_3| = \theta}) $$
$$ N_0 = 1 \times 10^{21} m^{-3} $$
$$ W_0 = |a_1| p_0^2 $$
$$ L_0 = p_0 \sqrt{\frac{a_0}{W_0}} = 3.9609 \times 10^{-8} m $$
$$ J_0 = G_0 L_0 $$
$$ \phi_0 = \frac{a_0 p_0}{L_0} $$.

The steady-state non-dimensionalized equations are

\begin{equation}
\frac{d^2\op_r}{d\ovr^2} - \frac{\partial \oW_{DL}}{\partial \op_r} - \frac{d\ophi}{d\ovr} = 0
\end{equation}
\begin{equation}
\frac{d^2\op_s}{d\ovr^2} - \frac{\partial \oW_{DL}}{\partial \op_s} = 0
\end{equation}
\begin{equation}
\overline{b}_0 \frac{d^2\ot_r}{d\ovr^2} - \frac{\partial \oW_{DL}}{\partial \ot_r} = 0
\end{equation}
\begin{equation}
\overline{b}_0 \frac{d^2\ot_s}{d\ovr^2} - \frac{\partial \oW_{DL}}{\partial \ot_s} = 0
\end{equation}
\begin{equation}
\oepsilon_0 \frac{d^2 \ophi}{d\ovr^2} - \frac{d\op_r}{d\ovr} + \delta (-\on_c + \op_v + \oN_d^+ - \oN_a^-) = 0
\end{equation}

\begin{equation}
\oJ_n = -K_n(\frac{d\on_c}{d\ovr} - \obeta \phi_0 \on_c \frac{d\ophi}{d\ovr})
\end{equation}
\begin{equation}
\oJ_p = -K_p(\frac{d\op_v}{d\ovr} + \obeta \phi_0 \op_v \frac{d\ophi}{d\ovr})
\end{equation}
\begin{equation}
-\frac{d}{d\ovr} \oJ_n + \oG - \oB(\on_c \op_v - \oN_i^2) = 0
\end{equation}
\begin{equation}
-\frac{d}{d\ovr} \oJ_p + \oG - \oB(\on_c \op_v - \oN_i^2) = 0,
\end{equation}

where

$$ \oepsilon_0 = \frac{\varepsilon_0 \phi_0}{L_0 p_0} $$
$$ \delta = \frac{q N_0 L_0}{p_0} $$
$$ K_n = \frac{\nu_n N_0}{q \beta L_0 J_0} $$
$$ K_p = \frac{\nu_p N_0}{q \beta L_0 J_0} $$
$$ \oB = \frac{B N_0^2}{G_0} $$.

\end{appendices}

%===================================================================================
%\section{References}
%===================================================================================
\bibliographystyle{unsrt}
\bibliography{References}

\end{document}